\documentclass[12pt]{iopart}
\usepackage{times}
\usepackage{graphicx}

\begin{document}

\title{Oases in the Desert: Three New Proposals}
\author{Ernest Ma}
\address{Physics Department, University of California, Riverside, USA}
\begin{abstract}
Two new nontrivial U(1) gauge symmetries are proposed,
one based on the particle content of the standard model and the other
on that of its supersymmetric generalization.
Each is an unexpected first example of its kind.  A third new proposal is the 
successful derivation of a realistic Majorana neutrino mass matrix, 
based on the underlying symmetry $A_4$ and its radiative corrections.
\end{abstract}

\section{Introduction}

The title of the talk I gave at {\it Beyond the Desert 2002} was ``New 
Unexpected Gauge Extensions of the Standard Model''.  At the request of 
the editor of these Proceedings, I am including some very recent work on 
the prediction of the Majorana neutrino mass matrix, hence the present title.

I start with a brief review of the symmetries of the Standard Model and 
previous (trivial) gauge extensions.  I then discuss two newly discovered 
nontrivial U(1) gauge extensions \cite{ma1,ma2}, based on the particle 
content of the Standard Model and its supersymmetric generalization 
respectively.  Finally I switch gear and present a new natural understanding 
of the Majorana neutrino mass matrix, which automatically gives the 
observed pattern of solar \cite{sol} and atmospheric \cite{atm} neutrino 
oscillations and a prediction of neutrinoless double beta decay at the level 
of 0.4 eV \cite{klapdor}.

\section{Anomaly Cancellation in the Standard Model}

The gauge group of the Standard Model is $SU(3)_C \times SU(2)_L \times 
U(1)_Y$, under which each family of quarks and leptons transforms as 
follows:
\begin{eqnarray}
&& (u,d)_L \sim (3,2,1/6), ~~ u_R \sim (3,1,2/3), ~~ d_R \sim (3,1,-1/3); \\ 
&& (\nu,e)_L \sim (1,2,-1/2), ~~ e_R \sim (1,1,-1).
\end{eqnarray}
The axial-vector triangle anomaly \cite{avv} is absent because \cite{bim}
\begin{eqnarray}
&& [SU(3)]^2~Y ~:~ \left( {1 \over 2} \right) \left[ 2 \left( {1 \over 6} 
\right) - {2 \over 3} - \left( -{1 \over 3} \right) \right] = 0, \\ 
&& [SU(2)]^2~Y ~:~ \left( {1 \over 2} \right) \left[ 3 \left( {1 \over 6} 
\right) + \left( -{1 \over 2} \right) \right] = 0, \\ 
&& Y^3 ~:~ 6 \left( {1 \over 6} \right)^3 - 3 \left( {2 \over 3} \right)^3 - 3 
\left( -{1 \over 3} \right)^3 + 2 \left( -{1 \over 2} \right)^3 - (-1)^3 = 0.
\end{eqnarray}
Note the nontrivial cancellation between quarks and leptons in the last two 
equations.  The original 1967 model of leptons by Weinberg \cite{wein67} was 
thus anomalous.

\section{Automatic Symmetries of the Standard Model}

Given the minimal particle content of the Standard Model, there are four 
automatic global symmetries: baryon number $B$ and the 3 lepton numbers 
$L_e$, $L_\mu$, $L_\tau$.  Each is anomalous but the combination 
$a_B B + a_e L_e + a_\mu L_\mu + a_\tau L_\tau$
is not for the following cases.\\

\noindent (1) If there is no $N_R \sim (1,1,0)$, then the only solution is 
$a_B=0$ and $(a_e, a_\mu, a_\tau) = (1,-1,0)$, $(1,0,-1)$, or $(0,1,-1)$.  
This allows $L_i - L_j$ to be gauged \cite{hjlv}.\\

\noindent (2) If only one family has an additional $N_R$, then the only 
solution is $a_B = 1$ and $(a_e, a_\mu, a_\tau) = (-3,0,0)$, $(0,-3,0)$, or 
$(0,0,-3)$.  This allows $B - 3L_i$ to be gauged \cite{msr}.\\

\noindent (3) If two $N_R$'s are added, then $a_B =1$ and $a_e=0$, $a_\mu +  
a_\tau = -3$, or $a_\mu=0$, $a_e + a_\tau = -3$, or $a_\tau=0$, $a_e + a_\mu 
= -3$ is a solution.  This allows for example $B - (3/2) (L_\mu + L_\tau)$ 
to be gauged \cite{mr}.\\

\noindent (4) If there are three $N_R$'s, then $a_B = 1$ and $a_e + a_\mu + 
a_\tau = -3$ is a solution.  The well-known example of $a_e = a_\mu 
= a_\tau = -1$ allows $B-L$ to be gauged \cite{mm}.  Another solution is 
$a_B=0$ and $a_e + a_\mu + a_\tau = 0$.  This means for example that 
$2L_e - L_\mu - L_\tau$ may be gauged.

\section{Neutrino Mass and a New U(1) Gauge Symmetry}

The addition of $N_R$ allows the doublet neutrinos $\nu_i$ to acquire 
small Majorana masses via the famous seesaw mechanism.  However, if $N_R$ 
is replaced by a heavy fermion triplet
\begin{equation}
(\Sigma^+, \Sigma^0, \Sigma^-)_R \sim (1,3,0),
\end{equation}
neutrinos get seesaw masses just as effectively \cite{flhj,ma98}.

Since the addition of one $N_R$ per family leads to the well-known U(1) gauge 
symmetry $B-L$, the same question may be raised as to the addition of one 
$\Sigma_R$ triplet per family.  It has recently been shown \cite{ma1} that 
indeed such an $U(1)_X$ exists, under which
\begin{eqnarray}
&& (u,d)_L \sim n_1, ~~ u_R \sim n_2 = {1 \over 4} (7n_1-3n_4), ~~ d_R \sim 
n_3 = {1 \over 4} (n_1 + 3n_4), \\ && (\nu, e)_L \sim n_4, ~~ e_R \sim n_5 = 
{1 \over 4} (-9n_1+5n_4), ~~ \Sigma_R \sim n_6 = {1 \over 4} (3n_1+n_4).
\end{eqnarray}
Note that this does not correspond to any combination of known quantum-number 
assignments, such as $Q$, $Y$, $B$ or $L$.

To show that $U(1)_X$ has no anomalies, consider first
\begin{eqnarray}
&& [SU(3)]^2~X ~:~ 2n_1-n_2-n_3 = 0,~~~{\rm and}~~~Y^2~X ~:~\\
&&  6 \left( {1 \over 6} \right)^2 n_1 - 3 \left( {2 \over 3} 
\right)^2 n_2 - 3 \left( -{1 \over 3} \right)^2 n_3 + 2 \left( -{1 \over 2} 
\right)^2 n_4 - (-1)^2 n_5 = 0,\\
&& {\rm and}~~~[SU(2)]^2~X ~:~ \left( {1 \over 2} \right) (3n_1+n_4) - 
(2) n_6 = 0.
\end{eqnarray}
These imply
\begin{equation}
n_3=2n_1-n_2, ~~~~ n_5 = -{1 \over 2} n_1 - n_2 + {1 \over 2} n_4, ~~~~ 
n_6 = {1 \over 4} (3n_1+n_4).
\end{equation}
\newpage
\noindent Consider next $~YX^2 ~:~$ 
\begin{equation}
6 \left( {1 \over 6} \right) n_1^2 - 3 \left( {2 \over 3} \right) 
n_2^2 - 3 \left( -{1 \over 3} \right) n_3^2 + 2 \left( -{1 \over 2} \right) 
n_4^2 - (-1) n_5^2 = 0.
\end{equation}
Using Eq.~(12), this implies
\begin{equation}
{1 \over 4} (3n_1+n_4) (7n_1-4n_2-3n_4) = 0.
\end{equation}
If $3n_1+n_4=0$, then $n_6=0$ as well, and $U(1)_X$ is proportional to 
$U(1)_Y$, i.e. no new gauge symmetry has been discovered.  On the other 
hand, if $7n_1-4n_2-3n_4=0$ is chosen instead, the solution given by Eqs.~(7) 
and (8) is obtained.  Nevertheless, there is still one more condition to be 
checked, i.e. the sum of the cubes of all $U(1)_X$ charges.  Remarkably,
\begin{equation}
X^3 ~:~ 6n_1^3 - 3n_2^3 - 3n_3^3 + 2n_4^3 - n_5^3 - 3n_6^3 = 0
\end{equation}
automatically.  Furthermore, the sum of all $U(1)_X$ charges themselves,
\begin{equation}
X ~:~ 6n_1 - 3n_2 - 3n_3 + 2n_4 - n_5 -3n_6 = 0
\end{equation}
as well.  Hence the mixed gravitational-gauge anomaly \cite{mixed} is also 
absent automatically. These are highly nontrivial results.

The Higgs sector of this gauge extension requires two doublets, one with 
$U(1)_X$ charge $(9n_1-n_4)/4$ to give mass to the charged leptons, and 
the other with charge $3(n_1-n_4)/4$ for the other fermions.
 
In general, for the fermion multiplet $(1,2p+1,0;n_6)_R$, there are 3 
conditions to be satisfied, i.e. the analogs of Eqs.~(11), (15) and (16):
\begin{eqnarray}
&& {1 \over 2} (3n_1+n_4) = {1 \over 3} p(p+1)(2p+1) n_6, \\
&& 6n_1^3 - 3n_2^3 - 3n_3^3 + 2n_4^3 - n_5^3 = (2p+1) n_6^3,\\
&& 6n_1 -3n_2 -3n_3 + 2n_4 - n_5 = (2p+1) n_6. 
\end{eqnarray}
For $p \neq 0$ and $3n_1+n_4 \neq 0$, these imply
\begin{equation}
{4n_6 \over 3n_1+n_4} = {6 \over p(p+1)(2p+1)} = {3 \over 2p+1} = \left( 
{3 \over 2p+1} \right)^{1/3}.
\end{equation}
This determines $p=1$ (or $p=-2$ which is the same as $p=1$ with $n_6 \to 
-n_6$ and $\Sigma_R \to \Sigma_L$.)  In other words, the solution I have 
found is nontrivial and unique.

\section{NuTeV Discrepancy}

Since the gauge boson $X$ couples to quarks and leptons according to 
Eqs.~(7) and (8), it may have measurable effects in precision electroweak 
data, such as the deep inelastic scattering of $\nu_\mu$ and $\bar \nu_\mu$ 
on nucleons in the NuTeV experiment \cite{nutev}.  A discrepancy has been 
reported in the value of the effective $\sin^2 \theta_W$, i.e. 
$0.2277 \pm 0.0013 \pm 0.0009$ versus the expected $0.2227 \pm 0.00037$ of 
the Standard Model.  To see how this may be explained in the $U(1)_X$ model, 
consider the effective Hamiltonian of neutrino interactions:
\begin{equation}
{\cal H}_{int} = {G_F \over \sqrt 2} \bar \nu \gamma^\mu (1-\gamma_5) \nu 
[\epsilon_L^q \bar q \gamma_\mu (1-\gamma_5)q + \epsilon_R^q \bar q \gamma_\mu 
(1+\gamma_5) q].
\end{equation}
Assume no $Z-X$ mixing (so that the precision $Z$-pole measurements are not 
affected), then
\begin{eqnarray}
&& \epsilon_L^u = {1 \over 2} - {2 \over 3} \sin^2 \theta_W + n_1 \zeta,\\ 
&& \epsilon_L^d = -{1 \over 2} + {1 \over 3} \sin^2 \theta_W + n_1 \zeta,\\ 
&& \epsilon_R^u = -{2 \over 3} \sin^2 \theta_W + n_2 \zeta,\\ 
&& \epsilon_R^d = {1 \over 3} \sin^2 \theta_W + n_3 \zeta,
\end{eqnarray}
where
\begin{equation}
\zeta = 2n_4 \left( {g_X^2 \over M_X^2} \right) \left( {M_Z^2 \over g_Z^2} 
\right).
\end{equation}
The NuTeV resluts versus the Standard Model predictions are:
\begin{eqnarray}
&& (\epsilon_L^u)^2 + (\epsilon_L^d)^2 = 0.3005 \pm 0.0014 ~~~{\rm versus} 
~~0.3042, \\ 
&& (\epsilon_R^u)^2 + (\epsilon_R^d)^2 = 0.0310 \pm 0.0011 ~~~{\rm versus} 
~~0.0301.
\end{eqnarray}
Consider $n_1=1$, $n_4=4/3$, then $n_2=3/4$ and $n_3=5/4$.  The discrepancies 
in the above are thus
\begin{eqnarray}
&& \Delta_L = -{2 \over 3} \sin^2 \theta_W \zeta + 2 \zeta^2, \\ 
&& \Delta_R = -{1 \over 6} \sin^2 \theta_W \zeta + {17 \over 8} \zeta^2.
\end{eqnarray}
Let $\zeta = \sin^2 \theta_W/6$, then a very good fit is obtained, i.e.
\begin{eqnarray}
&& \Delta_L = -0.0028 ~~~{\rm versus} ~~-0.0037 \pm 0.0014, \\ 
&& \Delta_R = +0.0016 ~~~{\rm versus} ~~+0.0009 \pm 0.0011.
\end{eqnarray}
This choice also implies that $G_X/G_F = \sin^2 \theta_W/16 = 0.014$. 
Note that this solution assumes that $X$ couples to $\mu$ and the $u$ 
and $d$ quarks.  If the same couplings were used for the electron, then 
the constraints from atomic parity violation would be grossly violated.

\section{New U(1) Gauge Extension of the Supersymmetric Standard Model}

In extending the Minimal Standard Model to include supersymmetry, one old
problem and two new problems have to be faced.  The old problem is neutrino 
mass.  This may be solved again by adding heavy singlet neutral superfields 
$N^c$ (analog of $N_R$).  The two new problems are rapid proton decay and 
the value of the $\mu$ term.  It has been shown recently \cite{ma2} that 
a nontrivial U(1) gauge extension exists which cures all three problems. 

Consider the following superfields with their $SU(3)_C \times SU(2)_L \times 
U(1)_Y \times U(1)_X$ assignments:
\begin{eqnarray}
&& Q = (u,d) \sim (3,2,1/6;n_1), ~~ u^c \sim (3^*,1,-2/3;n_2),\\ 
&& d^c \sim (3^*,1,1/3;n_3), ~~ L = (\nu,e) \sim (1,2,-1/2;n_4),\\ 
&& e^c \sim (1,1,1;n_5), ~~ N^c \sim (1,1,0;n_6),\\ 
&& \phi_1 \sim (1,2,-1/2;-n_1-n_3), ~~ \phi_2 \sim (1,2,1/2;-n_1-n_2).
\end{eqnarray}

\newpage
\noindent Without $U(1)_X$, the trilinear terms $LLe^c$, $LQd^c$, and 
$u^cd^cd^c$ are allowed in the superpotential, thus causing rapid proton 
decay.  The usual solution is to impose $R$ parity, i.e. $R \equiv 
(-1)^{3B+L+2j}$, to forbid these terms.  Even so, dimension-5 terms such 
as $\tilde q \tilde q q l$ are still allowed and may well be too big.
Secondly, the term $\mu \phi_1 \phi_2$ 
is allowed by supersymmetry, so there is no understanding of why $\mu$ 
should be of the order 1 TeV and not some very large mass such as the 
Planck scale or the string scale.

With $U(1)_X$, the $\mu$ problem is solved by replacing it by a singlet 
neutral superfield
\begin{equation}
\chi \sim (1,1,0;2n_1+n_2+n_3),
\end{equation}
where $2n_1+n_2+n_3 \neq 0$.  The subsequent spontaneous breaking of 
$U(1)_X$ at the TeV scale generates the desirable value of the effective 
$\mu$ parameter.  The solution of the proton decay problem now depends 
on finding an anomaly-free set of $n_i$'s which also forbids the undesirable 
terms mentioned above.  With the superfield content as it is, there is no 
solution.  However, if 2 sets of
\begin{equation}
U \sim (3,1,2/3;n_7), ~~ U^c \sim (3^*,1,-2/3;n_8),
\end{equation}
and 1 set of
\begin{equation}
D \sim (3,1,-1/3;n_7), ~~ D^c \sim (3^*,1,1/3;n_8)
\end{equation}
are added, a solution again appears.  This will involve the remarkable 
exact factorization of the sum of 11 cubic terms as shown below.

There are 8 $n_i$'s, but 3 conditions are imposed:
\begin{equation}
n_1+n_3 = n_4+n_5, ~~ n_1+n_2 = n_4 + n_6, ~~ n_7 + n_8 = -2n_1-n_2-n_3.
\end{equation}
These allow $\phi_1$ to give mass to the $d$ quarks and the charged leptons, 
$\phi_2$ to give mass to the $u$ quarks and the neutrinos (i.e. $\nu N^c$), 
and $\chi$ to give mass to the $U$ and $D$ quarks.  Consider now the 
anomaly-free conditions linear in $X$:
\begin{eqnarray}
&& X [SU(3)]^2 ~:~ 2n_1+n_2+n_3+n_7+n_8 = 0,\\
&& X [SU(2)]^2 ~:~ n_2+n_3 = 7n_1+3n_4,\\
&& X [Y]^2 ~:~ n_2 + n_3 = 7n_1+3n_4.
\end{eqnarray}
The first condition is automatically satisfied, while the other two 
conditions are identical and eliminate just one additional $n_i$.  
Let the remaining 4 independent $n_i$'s be $n_1$, $n_2$, $n_4$, and $n_7$, 
then $X^2 [Y]$ implies
\begin{eqnarray}
&& 3n_1^2 - 2n_2^2 + n_3^2 - n_4^2 + n_5^2 + 3n_7^2 - 3n_8^2 - (n_1+n_3)^2 + 
(n_1+n_2)^2 \nonumber \\ 
&& = 6(3n_1+n_4)(2n_1-4n_2-3n_7) = 0.
\end{eqnarray}
The condition $3n_1+n_4=0$ contradicts $2n_1+n_2+n_3 \neq 0$ and must be 
discarded.  Thus $2n_1-4n_2-3n_7=0$ is required and only 3 independent 
$n_i$'s are left, which are chosen finally to be $n_1$, $n_4$, and $n_6$. 
The most nontrivial condition, i.e. $X^3$, is then
\begin{eqnarray}
&& 18n_1^3 + 9n_2^3 + 9n_3^3 + 6n_4^3 + 3n_5^3 + 3n_6^3 + 9n_7^3 + 9n_8^3 
\nonumber \\ 
&& - 2(n_1+n_3)^3 - 2(n_1+n_2)^3 + (2n_1+n_2+n_3)^3 = 0.
\end{eqnarray}
Amazingly, this exactly factorizes to
\begin{equation}
-36(3n_1+n_4)(9n_1+n_4-2n_6)(6n_1-n_4-n_6) = 0.
\end{equation}
Whereas the first factor cannot be zero, either of the other two can be 
chosen to be zero, and two solutions have been found.  They are summarized 
in Table 1.

\newpage
\begin{table}[htb]
\begin{center}
{Table 1: Solutions (A) and (B) where $n_i = a n_1 + b n_4$.}\\
~\\
\begin{tabular}{||c|c|c||c|c||}
\hline \hline
& \multicolumn{2}{c||}{(A)} & \multicolumn{2}{c||}{(B)} \\ 
\cline{2-5}
 & $a$ & $b$ & $a$ & $b$ \\ 
\hline
$n_2$ & 7/2 & 3/2 & 5 & 0 \\ 
$n_3$ & 7/2 & 3/2 & 2 & 3 \\ 
$n_5$ & 9/2 & 1/2 & 3 & 2 \\ 
$n_6$ & 9/2 & 1/2 & 6 & --1 \\ 
$n_7$ & --4 & --2 & --6 & 0 \\ 
$n_8$ & --5 & --1 & --3 & --3 \\ 
\hline
$-n_1-n_3$ & --9/2 & --3/2 & --3 & --3 \\ 
$-n_1-n_2$ & --9/2 & --3/2 & --6 & 0 \\ 
$2n_1+n_2+n_3$ & 9 & 3 & 9 & 3 \\ 
\hline \hline
\end{tabular}
\end{center}
\end{table}

\begin{table}[htb]
\begin{center}
{Table 2: Conditions on $n_1$ and $n_4$ in (A) and (B).}\\
~\\
\begin{tabular}{||c|c||c|c|l||}
\hline \hline
\multicolumn{2}{||c||}{(A)} & \multicolumn{2} {c|}{(B)} & \\ 
\hline
$c$ & $d$ & $c$ & $d$ & $cn_1+dn_4 \neq 0$ forbids \\ 
\hline
3 & 1 & 3 & 1 & $\mu$ term \\ 
9 & 5 & 3 & 4 & $L$ violation \\ 
7 & 3 & 3 & 2 & $B$ violation \\ 
1 & 1 & 1 & 3 & $U^c$ as diquark \\ 
1 & 1 & 1 & 0 & $D^c$ as diquark \\ 
1 & 0 & 5 & --1 & $U$ as leptoquark \\ 
1 & 0 & 1 & 1 & $D$ as leptoquark \\ 
13 & 1 & 4 & 3 & $U^c$, $D^c$ as semiquarks \\ 
\hline \hline
\end{tabular}
\end{center}
\end{table}

\noindent In Table 2 the various conditions of forbidding $B$ and $L$ 
violation, etc. 
in the superpotential are listed.  The condition $3n_1+n_4 \neq 0$ which 
forbids the $\mu$ term also forbids the higher-dimensional terms 
$QQQL$ and $u^cu^cd^ce^c$, which are allowed by $R$ parity.  Thus proton 
decay is much more effectively suppressed in this case.  Note also that 
solutions (A) and (B) are the same for $n_1=n_4$.

\section{Naturally Small Dirac Neutrino Masses}

The sum of $X$ charges is
\begin{eqnarray}
&& 3(6n_1+3n_2+3n_3+2n_4+n_5+n_6) + 3(3n_7+3n_8) + 2(-n_1-n_3) \nonumber \\ 
&& +2(-n_1-n_2) + (2n_1+n_2+n_3) = 6(3n_1+n_4) \neq 0.
\end{eqnarray}
To get rid of this mixed gravitational-gauge anomaly, add the following 
singlet superfields in units of $3n_1+n_4$: one with charge 3, three ($S^c$) 
with charge $-2$, and three ($N$) with charge $-1$.  Then since
\begin{equation}
1(3) + 3(-2) + 3(-1) = -6, ~~~ 1(3)^3 + 3(-2)^3 + 3(-1)^3 = 0,
\end{equation}
the mixed gravitational-gauge anomaly is canceled without affecting Eq.~(45). 

\newpage
\noindent If now $n_6 = 3n_1+n_4$ is assumed, then there are 3 copies each of 
$N^c (n_6)$, $N (-n_6)$, and $S^c (-2n_6)$.  Thus $NN^c$ forms a large 
invariant mass $M$, $NS^c\chi$ implies a mass $m_2$ proportional to 
$\langle \chi \rangle$, and $\nu N^c \phi_2^0$ implies a mass $m_1$ 
proportional to $\langle \phi_2^0 \rangle$.  The $12 \times 12$ mass matrix 
spanning $(\nu,S^c,N,N^c)$ is then of the form
\begin{equation}
{\cal M} = \pmatrix{ 0 & 0 & 0 & m_1 \cr 0 & 0 & m_2 & 0 \cr 0 & m_2 & 0 & M 
\cr m_1 & 0 & M & 0}.
\end{equation}
With $m_1 \sim 10^2$ GeV, $m_2 \sim 10^3$ GeV, and $M \sim 10^{16}$ GeV, 
this implies that neutrinos have naturally small Dirac masses of order
\begin{equation}
m_\nu = {m_1 m_2 \over M} \sim 10^{-2}~{\rm eV}.
\end{equation}

\section{Nearly Degenerate Majorana Neutrino Masses}

I now switch gear and consider the derivation of the observed pattern of 
solar \cite{sol} and atmospheric \cite{atm} neutrino oscillations in the 
context of nearly degenerate Majorana neutrino masses.  This is based 
\cite{mrmm} on the non-Abelian discrete symmetry $A_4$ (i.e. the symmetry 
group of a regular tetrahedron, the simplest of the 5 perfect geometric 
solids).  It will be shown \cite{bmv} that radiative corrections automatically 
generate the desired neutrino mass matrix and if these come from softly 
broken supersymmetry, then the effective mass measured in neutrinoless 
double beta decay \cite{klapdor} should not be much smaller than about 0.4 eV.

Suppose that at some high energy scale, the charged lepton mass matrix and 
the Majorana neutrino mass matrix are such that after diagonalizing the 
former, i.e.
\begin{equation}
{\cal M}_l = \pmatrix {m_e & 0 & 0 \cr 0 & m_\mu & 0 \cr 0 & 0 & m_\tau},
\end{equation}
the latter is of the form
\begin{equation}
{\cal M}_\nu = \pmatrix {m_0 & 0 & 0 \cr 0 & 0 & m_0 \cr 0 & m_0 & 0}.
\end{equation}
From the high scale to the electroweak scale, one-loop radiative corrections 
will change ${\cal M}_\nu$ as follows:
\begin{equation}
({\cal M}_\nu)_{ij} \to ({\cal M}_\nu)_{ij} + R_{ik} ({\cal M}_\nu)_{kj} 
+ ({\cal M}_\nu)_{ik} R_{kj},
\end{equation}
where the radiative correction matrix is assumed to be of the most general 
form, i.e.
\begin{equation}
R = \pmatrix {r_{ee} & r_{e\mu} & r_{e\tau} \cr r_{e\mu}^* & r_{\mu\mu} & 
r_{\mu\tau} \cr r_{e\tau}^* & r_{\mu\tau}^* & r_{\tau\tau}}.
\end{equation}
Thus the observed neutrino mass matrix is given by
\begin{equation}
{\cal M}_\nu = m_0 \pmatrix {1+2r_{ee} & r_{e\tau} + r_{e\mu}^* & r_{e\mu} + 
r_{e\tau}^* \cr r_{e\mu}^* + r_{e\tau} & 2r_{\mu\tau} & 1+r_{\mu\mu}+
r_{\tau\tau} \cr r_{e\tau}^* + r_{e\mu} & 1+r_{\mu\mu}+r_{\tau\tau} & 
2r_{\mu\tau}^*}.
\end{equation}
Consider first the case where all the parameters are real.  Redefine them 
as follows:
\begin{eqnarray}
&& \delta_0 \equiv r_{\mu\mu} + r_{\tau\tau} - 2r_{\mu\tau}, \\ 
&& \delta \equiv 2r_{\mu\tau}, \\
&& \delta' \equiv r_{ee} - {1 \over 2} r_{\mu\mu} - {1 \over 2} r_{\tau\tau} 
- r_{\mu\tau}, \\
&& \delta'' \equiv r_{e\mu} + r_{e\tau}.
\end{eqnarray}
Then the neutrino mass matrix becomes
\begin{equation}
{\cal M}_\nu = m_0 \pmatrix{1+\delta_0+2\delta+2\delta' & \delta'' & \delta'' 
\cr \delta'' & \delta & 1+\delta_0+\delta \cr \delta'' & 1+\delta_0+\delta & 
\delta}.
\end{equation}
This matrix is $exactly$ diagonalized with the eigenvalues
\begin{eqnarray}
&& m_1 = m_0 (1+2\delta+\delta'-\sqrt{\delta'^2+2\delta''^2}), \\
&& m_2 = m_0 (1+2\delta+\delta'+\sqrt{\delta'^2+2\delta''^2}), \\
&& m_3 = -m_0,
\end{eqnarray}
where $\delta_0$ has been set equal to zero by a trivial rescaling of the 
other parameters, and the neutrino mixing matrix is given by
\begin{equation}
\pmatrix{\nu_e \cr \nu_\mu \cr \nu_\tau} = \pmatrix {\cos \theta & 
-\sin \theta & 0 \cr \sin \theta/\sqrt 2 & \cos \theta/\sqrt 2 & -1/\sqrt 2 
\cr \sin \theta/\sqrt 2 & \cos \theta/\sqrt 2 & 1/\sqrt 2} \pmatrix {\nu_1 
\cr \nu_2 \cr \nu_3},
\end{equation}
where
\begin{equation}
\tan \theta = {\sqrt 2 \delta'' \over \sqrt{\delta'^2+2\delta''^2} - \delta'}, 
~~~~(\delta' < 0).
\end{equation}
This is exactly the right description of present data on solar and 
atmospheric neutrino oscillations, with $\sin^2 2 \theta_{atm} =1$ and 
$\tan^2 \theta_{sol} = 0.38$ if $|\delta''/\delta'| = \sqrt 2$ for example. 
It also predicts that the common mass of the three neutrinos, i.e. $m_0$, 
is what is measured in neutrinoless double beta decay.

\section{Discrete $A_4$ Symmetry}

The successful derivation of Eq.~(64) depends on having Eqs.~(51) and (52). 
To be sensible theoretically, they should be maintained by a symmetry. 
At first sight, it appears impossible that there can be a symmetry which 
allows them to coexist.  Here is where the non-Abelian discrete symmetry 
$A_4$ comes into play.  The key is that $A_4$ has three inequivalent 
one-dimensional representations \underline {1}, \underline {1}$'$, 
\underline {1}$''$, and one three-dimensional reprsentation \underline {3}, 
with the decomposition
\begin{equation}
\underline {3} \times \underline {3} = \underline {1} + \underline {1}' + 
\underline {1}'' + \underline {3} + \underline {3}.
\end{equation}
This allows the following natural assignments of quarks and leptons:
\begin{eqnarray}
&& (u_i,d_i)_L, ~~ (\nu_i,e_i)_L \sim \underline {3}, \\ 
&& u_{1R}, ~~ d_{1R}, ~~ e_{1R} \sim \underline {1}, \\ 
&& u_{2R}, ~~ d_{2R}, ~~ e_{2R} \sim \underline {1}', \\ 
&& u_{3R}, ~~ d_{3R}, ~~ e_{3R} \sim \underline {1}''.
\end{eqnarray}
Heavy fermion singlets are then added:
\begin{equation}
U_{iL(R)}, ~~ D_{iL(R)}, ~~ E_{iL(R)}, ~~ N_{iR} \sim \underline {3},
\end{equation}
together with the usual Higgs doublet and new heavy singlets:
\begin{equation}
(\phi^+,\phi^0) \sim \underline {1}, ~~~~ \chi^0_i \sim \underline {3}.
\end{equation}
With this structure, charged leptons acquire an effective Yukawa coupling 
matrix $\bar e_{iL} e_{jR} \phi^0$ which has 3 arbitrary eigenvalues 
(because of the 3 independent couplings to the 3 inequivalent one-dimensional 
representations) and for the case of equal vacuum expectation values of 
$\chi_i$, i.e.
\begin{equation}
\langle \chi_1 \rangle = \langle \chi_2 \rangle = \langle \chi_3 \rangle = u,
\end{equation}
the unitary transformation $U_L$ which diagonalizes ${\cal M}_l$ is given by
\begin{equation}
U_L = {1 \over \sqrt 3} \pmatrix {1 & 1 & 1 \cr 1 & \omega & \omega^2 \cr 
1 & \omega^2 & \omega},
\end{equation}
where $\omega = e^{2\pi i/3}$.  This implies that the effective neutrino 
mass operator, i.e. $\nu_i \nu_j \phi^0 \phi^0$, is proportional to
\begin{equation}
U_L^T U_L = \pmatrix {1 & 0 & 0 \cr 0 & 0 & 1 \cr 0 & 1 & 0},
\end{equation}
exactly as desired \cite{mrmm,bmv}.

\section{Softly Broken Supersymmetry}

To derive Eq.~(75), the validity of Eq.~(73) has to be proved.  This is 
naturally accomplished in the context of supersymmetry.  Let $\hat \chi_i$ 
be superfields, then its superpotential is given by
\begin{equation}
\hat W = {1 \over 2} M_\chi (\hat \chi_1 \hat \chi_1 + \hat \chi_2 \hat \chi_2 
+ \hat \chi_3 \hat \chi_3) + h_\chi \hat \chi_1 \hat \chi_2 \hat \chi_3.
\end{equation}
Note that the $h_\chi$ term is invariant under $A_4$, a property not found 
in $SU(2)$ or $SU(3)$.  The resulting scalar potential is
\begin{equation}
V = |M_\chi + h_\chi \chi_2 \chi_3|^2 + |M_\chi \chi_2 + h_\chi \chi_3 
\chi_1|^2 + |M_\chi \chi_3 + h_\chi \chi_1 \chi_2|^2.
\end{equation}
Thus a supersymmetric vacuum $(V=0)$ exists for
\begin{equation}
\langle \chi_1 \rangle = \langle \chi_2 \rangle = \langle \chi_3 \rangle = u 
= -M_\chi /h_\chi,
\end{equation}
proving Eq.~(73), with the important additional result that the spontaneous 
breaking of $A_4$ at the high scale $u$ does not break the supersymmetry.

To generate the proper radiative corrections which will result in a 
realistic Majorana neutrino mass matrix, $A_4$ is assumed broken also 
by the soft supersymmetry breaking terms.  In particular, the mass-squared 
matrix of the left sleptons will be assumed to be arbitrary.  This allows 
$r_{\mu\tau}$ to be nonzero through $\tilde \mu_L - \tilde \tau_L$ mixing, 
from which the parameter $\delta$ may be evaluated.  For illustration, 
using the approximation that $\tilde m_1^2 >> \mu^2 >> M_{1,2}^2 = 
\tilde m_2^2$, where $\mu$ is the Higgsino mass and $M_{1,2}$ are gaugino 
masses, I find
\begin{equation}
\delta = {\sin \theta \cos \theta \over 16 \pi^2} \left[ (3g_2^2-g_1^2) \ln 
{\tilde m_1^2 \over \mu^2} - {1 \over 4} (3g_2^2+g_1^2) \left( \ln 
{\tilde m_1^2 \over \tilde m_2^2} - {1 \over 2} \right) \right].
\end{equation}
Using $\Delta m_{32}^2 = 2.5 \times 10^{-3}$ eV$^2$ from the atmospheric 
neutrino data, this implies that
\begin{equation}
\left[ \ln {\tilde m_1^2 \over \mu^2} - 0.3 \left( \ln {\tilde m_1^2 \over 
\tilde m_2^2} - {1 \over 2} \right) \right] \sin \theta \cos \theta 
\simeq 0.535 \left( {0.4 ~{\rm eV} \over m_0} \right)^2.
\end{equation}
To the extent that the factor on the left cannot be much greater than unity, 
this means that $m_0$ cannot be much smaller than about 0.4 eV \cite{klapdor}.

\section{Nonzero $U_{e3}$ and CP Violation}

The matrix of Eq.~(55) has complex phases.  It can easily been shown that 
only one phase remains after all possible redefinitions.  As a result, the 
most general Majorana neutrino mass matrix derivable from $A_4$ is actually of 
the form
\begin{equation}
{\cal M}_\nu = m_0 \pmatrix {1 + 2\delta + 2\delta' & \delta'' & \delta''^* 
\cr \delta'' & \delta & 1 + \delta \cr \delta''^* & 1 + \delta & \delta},
\end{equation}
where only $\delta''$ is complex.  In this case, $U_{e3}$ becomes nonzero, 
i.e.
\begin{equation}
U_{e3} \simeq {i Im \delta'' \over \sqrt 2 \delta}.
\end{equation}
Note the important result that $U_{e3}$ is purely imaginary.  Thus CP 
violation in neutrino oscillations is predicted to be maximal, which is 
desirable for future long-baseline neutrino experiments.\\

In the presence of $Im \delta''$, the previous expressions are still 
approximately valid with the replacement of $\delta'$ by $\delta' + 
(Im \delta'')^2 /2 \delta$ and of $\delta''$ by $Re \delta''$.  There 
is also the relationship
\begin{equation}
\left[ {\Delta m_{12}^2 \over \Delta m_{32}^2} \right]^2 \simeq \left[ 
{\delta' \over \delta} + |U_{e3}|^2 \right]^2 + \left[ {Re \delta'' \over 
\delta} \right]^2.
\end{equation}
Using $\Delta m_{12}^2 \simeq 5 \times 10^{-5}$ eV$^2$ from solar neutrino 
data and $|U_{e3}| < 0.16$ from reactor neutrino data \cite{react}, I find
\begin{eqnarray}
&& Im \delta'' < 8.8 \times 10^{-4} ~(0.4~{\rm eV}/m_0)^2, \\ 
&& Re \delta'' < 7.8 \times 10^{-5} ~(0.4~{\rm eV}/m_0)^2.
\end{eqnarray}

\section{Conclusions}

There are undoubtedly oases in the desert beyond the Standard Model.  Two 
first examples of nontrivial U(1) gauge extensions of the Standard Model 
and its supersymmetric generalization have been discovered.  In the first 
case \cite{ma1}, a heavy lepton triplet $(\Sigma^+, \Sigma^0, \Sigma^-)$ per 
family is added to the Standard Model as the anchor for a naturally small 
seesaw Majorana neutrino mass (instead of using the canonical singlet 
$N_R$). This allows for the remarkable existence of a new U(1) gauge 
symmetry, with possible phenomenological consequences already at the 
electroweak scale, such as the NuTeV discrepancy.  In the second case 
\cite{ma2}, with the addition of new superfields to the Supersymmetric 
Standard Model, a new U(1) gauge symmetry is again discovered which forbids 
proton decay, explains the size of the $\mu$ term, and allows for naturally 
small Dirac neutrino masses.  The existence of this new gauge symmetry 
involves the highly nontrivial exact factorization of the sum of 11 cubic 
terms.\\

\noindent A third possible oasis is that of an underlying $A_4$ symmetry at 
some high energy scale, which allows the observed Majorana neutrino mass 
matrix to be derived from radiative corrections.  It has been shown \cite{bmv} 
that this automatically leads to $\sin^2 2 \theta_{atm} = 1$ and a large 
(but not maximal) solar mixing angle.  Using neutrino oscillation data, 
and assuming radiative corrections from soft supersymmetry breaking, the 
effective mass measured in neutrinoless double beta decay is predicted to 
be not much less than 0.4 eV.

\section*{Acknowledgements}

I thank Hans Klapdor, Juha Peltoniemi, and the other organizers of Beyond 
2002 for their great hospitality at Oulu. 
This work was supported in part by the U.~S.~Department of Energy under 
Grant No.~DE-FG03-94ER40837.


\end{document}